\documentclass[sigconf]{acmart}

\AtBeginDocument{%
  }

\copyrightyear{2026}
\acmYear{2026}
\setcopyright{cc}
\setcctype{by}
\acmConference[DFF '26]{2nd Deepfake Forensics Workshop: Detection, Attribution, Recognition, and Adversarial Challenges in the Era of AI-Generated Media}{November 10--14, 2026}{Rio de Janeiro, Brazil}
\acmBooktitle{2nd Deepfake Forensics Workshop: Detection, Attribution, Recognition, and Adversarial Challenges in the Era of AI-Generated Media (DFF '26), November 10--14, 2026, Rio de Janeiro, Brazil}
\acmDOI{10.1145/3840473.3840518}
\acmISBN{979-8-4007-2927-0/2026/11}

\usepackage{amsmath,amssymb,amsfonts}
\usepackage{algorithmic}
\usepackage{graphicx}
\usepackage{textcomp}
\usepackage{xcolor}
\usepackage{glossaries}
\usepackage{url}
\usepackage{siunitx}
\usepackage{booktabs, siunitx, tabularx, stfloats}
\usepackage{hyperref}
\usepackage{cleveref}
\usepackage{enumitem}
\usepackage{bm}
\usepackage{comment}
\usepackage{subcaption}
\usepackage{flushend}
\usepackage{subcaption}

\crefname{section}{Section}{Sections}
\crefname{figure}{Figure}{Figures}
\crefname{table}{Table}{Tables}

\newacronym{tts}{TTS}{Text to Speech}
\newacronym{lcnn}{LCNN}{Light-CNN}
\newacronym{roc}{ROC}{Receiver Operating Characteristic}
\newacronym{moe}{MoE}{Mixture of Experts}
\newacronym{eer}{EER}{Equal Error Rate}
\newacronym{auc}{AUC}{Area Under the Curve}
\newacronym{mlp}{MLP}{Multi-Layer Perceptron}
\newacronym{llr}{LLR}{Log-Likelihood Ratio}
\newacronym{kde}{KDE}{Kernel Density Estimation}
\newacronym{lsf}{LSF}{Line Spectral Frequencies}
\newacronym{hmm}{HMM}{Hidden Markov Model}
\newacronym{hts}{HTS}{HMM-based TTS}
\newacronym{stft}{STFT}{Short-Time Fourier Transform}
\newacronym{cllr}{Cllr}{Log-Likelihood Ratio Cost}
\newacronym{dl}{DL}{Deep Learning}
\newacronym{balr}{BA-LR}{Binary-attribute-based-LR}

\begin{document}

\title{Toward Interpretable Speech Deepfake Detection using Artifact-Specific Experts and Calibrated Detection Scores}

\author{Viola Negroni}
\email{viola.negroni@polimi.it}
\orcid{0009-0000-2483-4366}
\affiliation{%
  \institution{DEIB, Politecnico di Milano}
  \city{Milano}
  \country{Italy}
}
\author{Xin Wang}
\email{wangxin@nii.ac.jp}
\orcid{}
\affiliation{%
  \institution{National Institute of Informatics}
  \city{Tokyo}
  \country{Japan}
}
\author{Wanying Ge}
\email{gewanying@nii.ac.jp}
\orcid{}
\affiliation{%
  \institution{National Institute of Informatics}
  \city{Tokyo}
  \country{Japan}
}
\author{Paolo Bestagini}
\email{paolo.bestagini@polimi.it}
\orcid{0000-0003-0406-0222}
\affiliation{%
  \institution{DEIB, Politecnico di Milano}
  \city{Milano}
  \country{Italy}
}
\author{Junichi Yamagishi}
\email{jyamagis@nii.ac.jp}
\orcid{}
\affiliation{%
  \institution{National Institute of Informatics}
  \city{Tokyo}
  \country{Japan}
}
\author{Stefano Tubaro}
\email{stefano.tubaro@polimi.it}
\orcid{0000-0002-1990-9869}
\affiliation{%
  \institution{DEIB, Politecnico di Milano}
  \city{Milano}
  \country{Italy}
}

\renewcommand{\shortauthors}{Negroni et al.}

\begin{abstract}
    In this work, we propose an interpretable framework for speech deepfake detection based on artifact-specific expert models. 
    Rather than relying on black-box decisions, the framework provides human-understandable evidence, which is critical in high-stakes settings.
    Each expert is trained to detect a specific speech synthesis artifact, and its output is calibrated into a log-likelihood ratio that serves as an interpretable evidence score.
    We evaluate five artifact-specific experts and show that, with proper calibration, they can capture their target artifacts and produce meaningful evidence. 
    Importantly, each expert estimates only the presence of its assigned artifact rather than directly performing the final decision. 
    Their outputs are aggregated into an ensemble to produce the actual real-versus-fake classification, while maintaining interpretability by indicating how strongly each expert supports or contradicts a fake classification.
    Results show that artifact-specific experts capture interpretable signals of synthetic speech across multiple generation pipelines. 
\end{abstract}





\maketitle

\section{Introduction}
\label{sec:intro}

Speech deepfake detection aims to identify synthetic speech used for malicious purposes, such as impersonation, fraud, and misinformation.
Over recent years, substantial progress has been achieved in detection performance~\cite{jung2022aasist,huang2025_speechfake,xiao2024xlsr-mamba}. This progress has coincided with a shift from simpler machine learning approaches to increasingly complex \gls{dl}-based systems, ranging from supervised models~\cite{tak2021end, jung2022aasist} to large pre-trained models~\cite{wang2022investigating, antideepfake_2025} and, more recently, large language model-based methods~\cite{gu2025allm4add,xie2026ftgrpo}.
In parallel, manually designed features~\cite{borrelli2021synthetic,todisco2017cqcc} have progressively been replaced by learned self-supervised representations~\cite{salvi2024comparative}. 
While these developments have improved performance, they have also reduced interpretability compared to classical approaches (e.g., decision trees and phonetic features~\cite{machado_towards_2025}).

In high-stakes applications such as deepfake detection, where decisions may carry forensic, legal, or societal consequences, this lack of transparency becomes particularly problematic. 
Some post-hoc methods have been proposed to explain the behavior of black-box detectors~\cite{geExplainable2022a}. 
However, relying on post-hoc explanations for black-box models is fundamentally limited~\cite{rudin2019stop}, since these explanations do not guarantee faithful correspondence with the actual reasoning process of the classifier.
Instead, interpretability should be embedded directly into the model design, so that the explanation itself corresponds to the mechanism used to reach the decision.

Recent work has explored related directions. 
One approach~\cite{mishra2026towards} augments real/fake detectors with auxiliary classifiers predicting attributes of the synthesis process, such as TTS vs. voice conversion and the generator’s DNN type. Another~\cite{negroni2026multitask} adds branches predicting acoustic features and per-frame weights.
While these methods provide additional signals, the resulting explanations remain coarse and do not directly expose nor quantify the evidence driving individual detection decisions.

\begin{figure}
  \includegraphics[width=0.99\columnwidth]{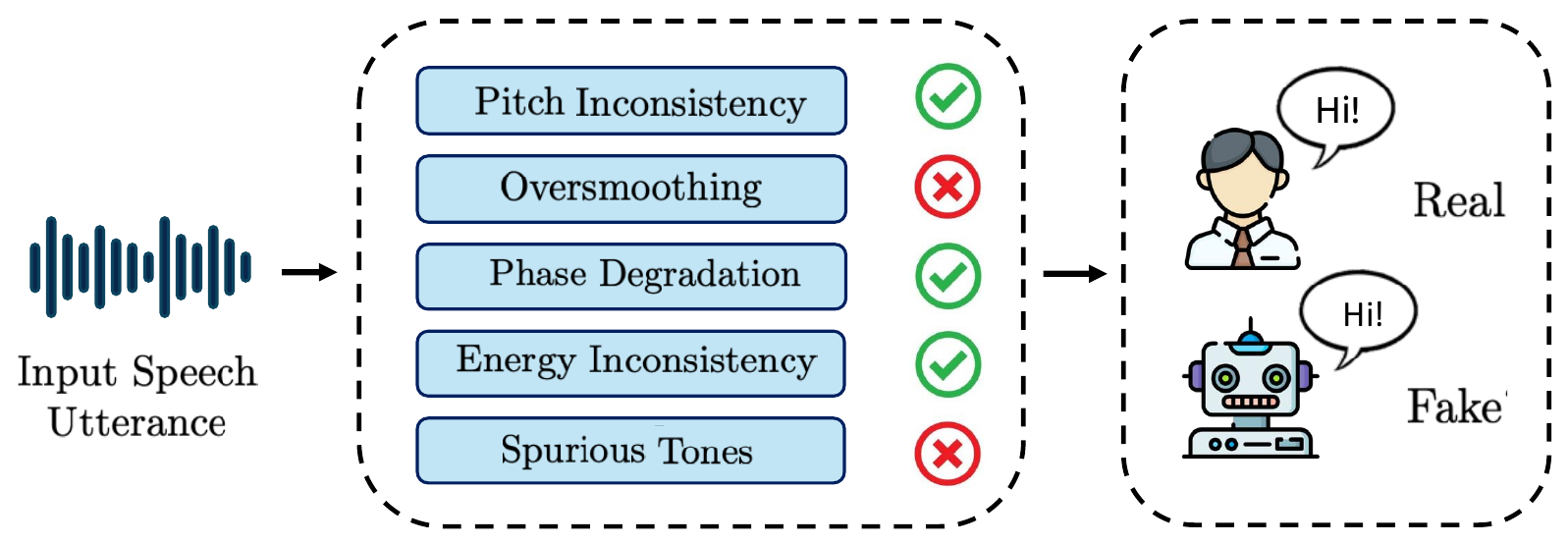}
  \vspace{-2mm}
  \caption{Overview of the proposed framework for interpretable speech deepfake detection.}
  \label{fig:overview}
\end{figure}

Our proposed framework explicitly targets intrinsically interpretable detection. 
As shown in \Cref{fig:overview}, the task is decomposed into predefined \textit{synthesis artifacts}, each handled by a dedicated expert model.
Expert outputs are calibrated into interpretable \glspl{llr}, quantifying evidence for the presence of each artifact, and then aggregated into a global score. 
To the best of our knowledge, this is the first work to frame speech deepfake detection as explicit artifact identification followed by \gls{llr}-based aggregation.

Conventional \gls{dl}-based models implicitly learn artifacts without defining or isolating them, making it unclear whether decisions rely on meaningful cues or spurious correlations~\cite{sahidullah2025shortcut}.
This raises concerns about their reliability. 
In contrast, our approach is expected to mitigate the issue by replacing implicit artifact learning with an explicit formulation, where predefined synthesis artifacts are detected individually, and their evidence is combined via a principled \gls{llr}-based framework. 
Relatedly, the \gls{balr} framework for speaker verification~\cite{amor_balr_2023} also produces attribute-specific LLRs, but it decomposes scores from a pretrained verifier using a surrogate model. 
This top-down approach differs from ours, which directly computes and aggregates individual \glspl{llr} in a bottom-up manner.

As this is a preliminary, exploratory study, we focus on a set of artifacts well understood in the speech synthesis field. 
Experimental results show that the proposed method achieves meaningful detection performance while enabling the contribution of each expert to the final decision to be directly quantified and interpreted in terms of the presence of each artifact. 
Additionally, the framework is modular by design, and new experts can be directly incorporated. 
This additive design is well-suited to real-world deployment and may also facilitate the identification of dataset shortcuts and spurious correlations (see \S~\ref{subsec:res_5}).

\section{Proposed Method}

\label{sec:method}
\begin{figure*}
  \includegraphics[width=.8\textwidth]{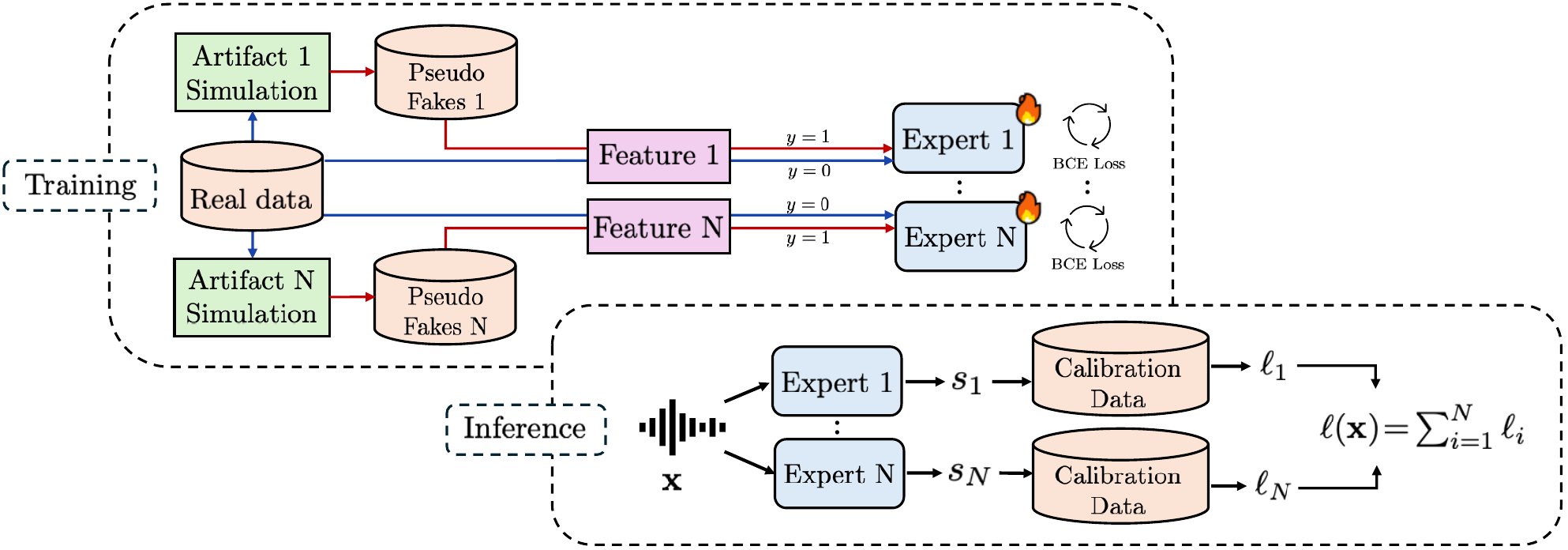}
  \vspace{-3mm}
  \caption{Detailed representation of the proposed framework (training and inference).}
  \label{fig:detailed}
\end{figure*}

In this section, we describe the proposed interpretable speech deepfake detection framework (detailed representation in \Cref{fig:detailed}) in the order of the problem formulation (\S~\ref{subsec:problem}), the design of artifact-specific expert models (\S~\ref{subsec:met_1}), their calibration into \glspl{llr} (\S~\ref{subsec:met_2}), and the aggregation of \glspl{llr} (\S~\ref{subsec:met_3}).

\subsection{Problem Formulation}
\label{subsec:problem}
Given a discrete speech signal $\mathbf{x}$ with class label $y \in \{0,1\}$, we consider two hypotheses:
$H_0$ denotes the real class ($y = 0$) where $\mathbf{x}$ is free of artifacts; and $H_1$ denotes the fake class ($y = 1$) where $\mathbf{x}$ contains synthesis artifacts.
The task is to determine the class of $\mathbf{x}$ using multiple artifact-specific experts.
Each expert $\mathcal{E}_i(\cdot)$ targets a particular synthesis artifact and uses $\mathcal{F}_i$ to extract features relevant to the detection of the artifact. The raw output score of each expert $s_i\triangleq \mathcal{E}_i(\mathcal{F}_i(\mathbf{x}))$ is then calibrated towards a \gls{llr} using a calibration function $\mathcal{C}$, which can be written as
\begin{equation}\mathcal{C}(s_i)\approx \ell_i(\mathbf{x}) \triangleq \log \frac{p(\mathcal{F}_i(\mathbf{x}) \mid H_1)}{p(\mathcal{F}_i(\mathbf{x}) \mid H_0)}.
\end{equation}

An ideal LLR quantifies the evidence in favor of artifact presence, which is the foundation for ensembling multiple experts. 
An ensemble with $I$ experts can produce a system-level LLR $\ell(\mathbf{x})$ by
\begin{align}
\ell(\mathbf{x})\triangleq\log\frac{p(\mathbf{x} | H_1)}{p(\mathbf{x} | H_0)} &\approx \log\frac{p(\{\mathcal{F}_1(\mathbf{x}),\cdots, \mathcal{F}_I(\mathbf{x})\} | H_1)}{p(\{\mathcal{F}_1(\mathbf{x}),\cdots, \mathcal{F}_I(\mathbf{x})\} | H_0)} \label{eq:approx_artifacts}\\
&\approx \sum_{i=1}^{I}\log\frac{p(\mathcal{F}_i(\mathbf{x}) | H_1)}{p(\mathcal{F}_i(\mathbf{x}) | H_0)} \label{eq:llr_independence}.
\end{align}
The first approximation (Eq.~\eqref{eq:approx_artifacts}) assumes that the set of artifacts $\{\mathcal{F}_1(\mathbf{x}),\cdots, \mathcal{F}_I(\mathbf{x})\}$ summarizes all the necessary artifacts to make the decision. The second approximation (Eq.~\eqref{eq:llr_independence}) assumes independence of the feature distributions given the hypothesis ($H_0$ or $H_1$), which is known as the \emph{independent additivity} of individual LLRs~\cite{jaynes2003probability} and is widely used in multi-factor biometric authentication~\cite{alonso2022, bassit_fast_2021}.\footnote{The above formulation also differ from \gls{balr}~\cite{amor_balr_2023}, which solves the problem of decomposing $\ell(\mathbf{x})$ into $\{\ell_i(\mathbf{x})\triangleq \log \frac{p(\mathcal{F}_i(\mathbf{x}) \mid H_1)}{p(\mathcal{F}_i(\mathbf{x}) \mid H_0)}\}_{i=1}^{I}$ and learning the mapping from each $\ell_i(\mathbf{x})$ to a specific speaker attribute.}
In the context of deepfake detection, if each expert focuses on a different artifact and captures a different aspect of the signal, 
the independent additivity of LLRs is a practical approximation.
Given $\ell(\mathbf{x})$, the ensemble system can decide $H_0$ and $H_1$ by comparing it with a pre-defined threshold following the Bayes decision theory.\footnote{Setting the threshold based on priors is well discussed in other literature~\cite[Eq.~(6)]{van2007introduction}. The threshold used in this study is discussed in \S~\ref{sec:setup}} 

Several technical questions need to be addressed.
First, we need to design the experts and their corresponding $\{\mathcal{F}_i\}$. Second, we need to properly implement the calibration function $\mathcal{C}$ so that $\mathcal{C}(s_i)\approx \ell_i(\mathbf{x})$.
Finally, since LLRs are assumed to be independent, we may try alternative ways to fuse the LLRs.

\subsection{Expert Model Design}
\label{subsec:met_1}
Taking advantage of the knowledge from the speech synthesis field, we design experts and their corresponding $\{\mathcal{F}_i\}$ for a set of well-known artifacts. 
An artifact might be regarded as \textit{a systematic and detectable deviation from real speech}, potentially arising at different stages of synthesis (e.g., acoustic modeling or waveform generation).
For each type of artifact, we construct an expert $\mathcal{E}_i(\cdot)$ that targets its detection. 
The resulting set of experts forms the building block of our method, allowing decisions to be interpreted in terms of the presence or absence of specific artifacts in the given speech.

However, it is not straightforward to learn the parameters of an artifact-specific $\mathcal{E}_i(\cdot)$ from fake data that is synthesized by existing speech generative models, because they may carry compounded artifacts.
Moreover, training on real and fake samples offers limited control over the learned discriminative cues, and the model may end up relying on dataset-specific biases unrelated to the task~\cite{sahidullah2025shortcut}.
To address this issue, we train each expert using a pseudo-fake generation strategy, in which the target artifact is simulated through a controlled manipulation applied to real speech.
This ensures that the simulated artifact is the only difference between real and pseudo-fake data, encouraging each expert to focus on the intended artifact.

Specifically, for a given artifact, we create manipulated speech from real utterances such that the applied manipulation emulates the target artifact while leaving other characteristics unchanged. To further enforce specialization, the input feature representation is deliberately designed to emphasize the artifact of interest and suppress unrelated information.
Formally, for a given $\mathbf{x}$, we derive two training inputs for $\mathcal{E}_i(\cdot)$ using the artifact-specific feature extraction $\mathcal{F}_i(\cdot)$ and manipulation $\mathcal{M}_i(\cdot)$:
$\mathbf{z}_0 = \mathcal{F}_i(\mathbf{x})$ and $\mathbf{z}_1 = \mathcal{F}_i(\mathcal{M}_i(\mathbf{x}))$, with labels $y=0$ and $y=1$, respectively. Each expert $\mathcal{E}_i(\cdot)$ is then trained as a binary classifier using multiple pairs of $\mathbf{z}_0$ and $\mathbf{z}_1$. 
Using pseudo-fake data not only facilitates the learning of artifact-specific experts but also simplifies data collection. 
The set of artifacts investigated in this study is explained in \S~\ref{sec:artifacts}.

\subsection{Model Calibration}
\label{subsec:met_2}

\subsubsection{Motivation}
As mentioned in \S~\ref{subsec:problem}, we calibrate $s_i$ from each expert so that its calibrated version better approximates $\ell_i(\mathbf{x})$. In fact, using LLRs is not only preferred when fusing multiple experts (Eq.~\eqref{eq:llr_independence}) but also widely recognized by the forensic community as the proper way of representing ``the value of the evidence $\dots$ to be reported to a court of law''~\cite[\S~14.3]{ramos2017biometric}. 
To briefly explain the reasons, we note that making a Bayes decision given a single expert requires a comparison between the posterior odds $\log\frac{P(H_1|\mathcal{F}_i(\mathbf{x}))}{P(H_0|\mathcal{F}_i(\mathbf{x}))}$ against the decision cost ratio $\log\frac{C_0}{C_1}$, where $C_0$ and $C_1$ are the costs for falsely classifying $H_1$ and $H_0$, respectively.
Using Bayes’ rule, we get an equivalent  decision strategy, which can be written as
\begin{equation}
    \ell_i(\mathbf{x})\triangleq\log\frac{p(\mathcal{F}_i(\mathbf{x})|H_1)}{p(\mathcal{F}_i(\mathbf{x})|H_0)} \underset{H_0}{\overset{H_1}{\gtrless}} \log\frac{C_0}{C_1}+\log\frac{P(H_0)}{P(H_1)},
    \label{eq:bayes}
\end{equation}
where $P(H_0)$ and $P(H_1)$ denote the class priors.
The main reason for producing and reporting LLRs is that they provide a \textit{universal interpretation} of the strength of evidence in favor of a hypothesis~\cite{ramos2017biometric}. 
This interpretation is independent of the data prior probability, the backbone model, or input representation: for instance, $\ell_i(\mathbf{x})=0$ does not favor either one of the hypotheses, and $\ell_i(\mathbf{x})=\log 3$ means that ``the evidence supports $H_1$ against $H_0$ with a degree of 3 versus 1''.
\footnote{According to the guideline of the European network of forensic science institutes~\cite{champod2016enfsi}, we may also verbally explain the support of evidence represented by the LLR. 
For example, $\ell_i(\mathbf{x})>\log 10^6$ denotes \emph{extremely strong} support to $H_1$ while $\log 10^4>\ell_i(\mathbf{x})>\log 10^3$ denotes \emph{moderately strong}.} 
\textit{Universal interpretation} also means that the LLRs of different experts are on the same scale and can be directly summed together (Eq.~\eqref{eq:llr_independence}). 
A further advantage, as shown in Eq.~\eqref{eq:bayes}, is the separation between modeling and decision-making: the forensic examiner produces the LLR (left-hand side of Eq.~\eqref{eq:bayes}), while the decision threshold (right-hand side), based on priors and costs, is set by the court~\cite{ramos2017biometric}.

A binary classifier does not necessarily produce LLRs. 
For instance, a probability-like $s_i \in [0,1]$ cannot be interpreted as a log-likelihood ratio $\ell_i(\mathbf{z}) \in (-\infty,\infty)$. 
Even if transformed into an unbounded score, for example, using $f(x)=\log\frac{x}{1-x}$ for $x\in(0,1)$, the score still may not satisfy the statistical properties of true LLRs (e.g., ``the LLR of the LLR is the LLR''~\cite{mandasari2018speaker}).

\subsubsection{Methods}
There are many possible ways of calibrating $s_i$ so that $\mathcal{C}(s_i)\approx\ell_i(\mathbf{x})$. 
We consider a generative approach~\cite{VanLeeuwen2014}, which treats the expert outputs $s_i$ as samples generated from class-conditional scores distributions $\{\tilde{p}(s_i | H_1;\theta_1), \tilde{p}(s_i | H_0;\theta_0)\}$: after estimating the two distributions, we calibrate by $\mathcal{C}(s_i)=\log \frac{\tilde{p}(s_i | H_1;\theta_1)}{\tilde{p}(s_i | H_0;\theta_0)}$.

In practice, we need to choose the parametric form of such distributions. 
A fully flexible approach models each class distribution free-form, e.g., via \gls{kde}~\cite{cuccovillo2023calibration}.
Scores $s_i$ are first mapped via Platt scaling (i.e., logistic regression $\frac{1}{1+\exp(as_i+b)}$ with learned parameters $\{a,b\}$)~\cite{platt1999probabilistic}, ensuring a bounded and monotonic score space suitable for density estimation. 
After that, class-conditional densities are then estimated using Gaussian \gls{kde}.
While the \gls{kde}-based method is flexible, it provides no guarantee that $\mathcal{C}(s_i)$ is monotonic, which is often desirable for calibration.

A more constrained form assumes that $\tilde{p}(s_i | H_1;\theta_1)$ and $\tilde{p}(s_i | H_0;\theta_0)$ are Gaussian distributions with a tied standard deviation~\cite{vanleeuwen2013calibrated_lr}, i.e.,
\begin{equation}
    p(s_i | H_1) = \mathcal{N}(s_i|\mu_f, \sigma), \quad p(s_i | H_0) = \mathcal{N}(s_i|\mu_r, \sigma), 
\end{equation}
where the learnable parameters are $\theta_1 = \{\mu_f, \sigma\}$ and $\theta_0 = \{\mu_r, \sigma\}$. 
Under this assumption, the resulting calibration reduces to an affine transformation $\mathcal{C}(s_i)=a s_i + b$, where
\begin{equation}
    a = \frac{\mu_{\text{f}} - \mu_{\text{r}}}{\sigma^2}, \quad
    b = -\frac{a(\mu_{\text{f}} + \mu_{\text{r}})}{2}.
\end{equation}
This method is known as the constrained, maximum-likelihood, Gaussian (CMLG) calibration~\cite{vanleeuwen2013calibrated_lr}.

In this work, we experiment with KDE and CMLG-based approaches. 
Note that each method requires a calibration set that is as representative as possible of the target (test) distribution, so that the estimated densities and resulting \glspl{llr} are meaningful. 
If this set is not representative, the learned densities may be mismatched, leading to unreliable \glspl{llr}.
The choice of the calibration set is described in \S~\ref{sec:setup}.

\subsection{Ensemble Aggregation Strategy}
\label{subsec:met_3}
Once the \glspl{llr} are obtained, the straightforward approach is to sum them, motivated by the independent additivity of \glspl{llr} (Eq.~\eqref{eq:llr_independence}). 
We also consider two alternative strategies.
The first uses a max operator
\begin{equation}
\ell(\mathbf{x})=\max_i \ell_i(\mathbf{x}),
\end{equation}
which retains only the strongest expert response for each sample and assumes that a single artifact may be sufficient for detection.
The second follows a Mixture-of-Experts formulation~\cite{negroniLeveraging2024} and applies a learned gating mechanism
\begin{equation}
    \ell(\mathbf{x})=\sum_i w_i(\mathbf{x}) \ell_i(\mathbf{x}),
    \label{eq:gating}
\end{equation}
where sample-dependent weights ${w_i(\mathbf{x})}$ adaptively combine expert \glspl{llr} by emphasizing the most relevant ones. 
In practice, the calibrated per-expert scores are passed to a linear gating network with dimension equal to the number of experts, which outputs softmax-normalized weights $\{w_i(\mathbf{x})\}$.

\begin{table*}
\centering
\caption{Five artifacts investigated in this study. Abbreviations are defined for statistical parametric speech synthesis (SPSS) and line spectral frequency (LSF). Numbers in brackets for MLP-based expert architectures denote layer sizes.}
\label{tab:expert_summary}
\resizebox{\textwidth}{!}{
\setlength{\tabcolsep}{3pt}{
\renewcommand{\arraystretch}{1.2}
\begin{tabular}{l l l l l l}
    \toprule
    Artifact & Example of target algorithm or module  & Raw features & Expert input (\#. proxy features) & Manipulation & Expert Architecture \\
    \midrule
    $F_0$ inconsistencies   & Unit selection                          & $F_0$ trajectory          & variability, 1st/2nd-order dynamics, ratios (10)          & shifting, shuffling, compression & MLP (10–32–16–1) \\
    Oversmoothed spectra           & Acoustic models in SPSS                  & LSFs                      & spatial/temporal spacing dynamics, coeff. variance (13)   & gaussian smoothing               & MLP (13–32–16–1) \\
    Degraded phase                   & Vocoders w/o modeling phase                         & Unwrapped phase           & ~~~~~~~~~~~~~~~~~~~~-                                                         & minimum-phase reconstruction     & Light CNN (LCNN) \\
    Energy inconsistencies  & Unit selection \& SPSS          & Log-power spec.      & global + band energy, temporal deltas, drop stats (24)    & local attenuations               & MLP (24–32–16–1) \\
    Tonal artifacts         & DNN-based vocoders                & Spectral fingerprint      & ~~~~~~~~~~~~~~~~~~~~-                                                         & inject resonances                & MLP (225–128–32–1)\\
    \bottomrule
\end{tabular}}}
\vspace{-1em}
\end{table*}

\begin{figure*}[t]
  \centering
  \includegraphics[height=3.5cm]{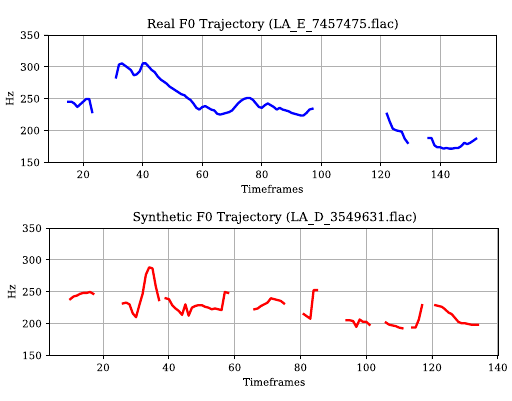}%
  \includegraphics[height=3.5cm]{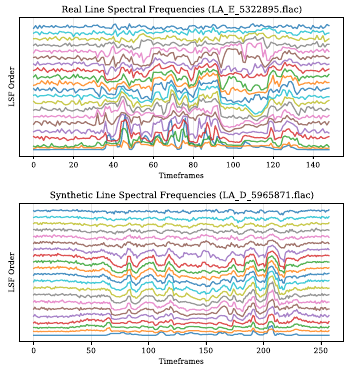}%
  \includegraphics[height=3.5cm]{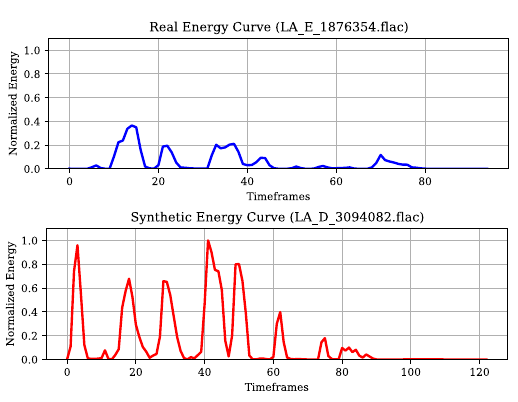}%
  \includegraphics[height=3.5cm]{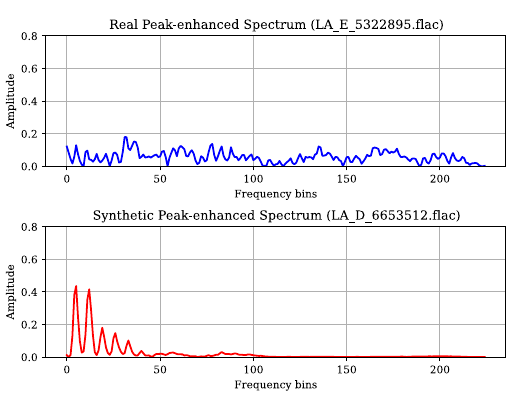}
  \caption{Illustrative real (top) vs.\ synthetic (bottom) examples for four
  of the targeted synthesis artifacts (left to right: $F_0$ inconsistency, oversmoothing,
  energy inconsistency, tonal artifact).}
  \label{fig:artifacts-examples}
\end{figure*}

\section{Artifacts and Expert Models}
\label{sec:artifacts}
This section outlines the artifacts and implementation of expert models for speech deepfake detection, designed based on our prior knowledge of speech synthesis algorithms.

Considering the variety of available approaches, we limit this preliminary work to \gls{tts}, the most widely used and extensively studied speech synthesis paradigm, and its generated artifacts. Specifically, we consider \num{5} major synthesis artifacts, spanning synthesis paradigms from unit selection to statistical parametric approaches, from linear models to neural methods; their implementation and association with specific deepfake families are detailed for each expert below (summary in \Cref{tab:expert_summary}). 

\subsection{F0 Inconsistencies}
\label{subsubsec:exp_1}
Generating a fundamental frequency ($F_0$) trajectory is essential for natural-sounding speech synthesis~\cite{hirschbergCommunication2002}. A known artifact in unit selection TTS is $F_0$ inconsistency~\cite{borilEffect2017}, perceived as prosodic jumps or discontinuities in synthetic speech. This arises from the concatenation of segmented units. 

We extract the $F_0$ trajectory using \texttt{pYIN}~\cite{mauch2014pyin}, and apply manipulations within voiced regions. The contour is segmented into 80–160 ms units, and segments are randomly selected with probability $p_{\text{seg}} \in [0.4, 0.7]$, scaled by factors in $[0.87, 0.92]$ or $[1.08, 1.30]$. Selected segments are partially shuffled with probability $p_{\text{shuffle}} \in [0.2, 0.5]$. Alternatively, with fixed probability, the $F_0$ contour is compressed toward its median (strength 0.4) and locally smoothed using a three-frame average (strength 0.4), reducing pitch variability. The procedure produces trajectories with either local discontinuities or reduced dynamics. Parameter ranges were empirically set to yield perceptually plausible yet pronounced artifacts for training.
We aggregate $F_0$ contours into scalar descriptors, forming a 10-dimensional feature vector for a \gls{mlp} classifier. Direct use of raw $F_0$ contours was ineffective.

\subsection{Oversmoothing}
\label{subsubsec:exp_2}
In contrast to unit selection, statistical parametric speech synthesis (SPSS) uses statistical acoustic models to generate smoothed $F_0$ and spectral trajectories~\cite{tokuda2013speech}. 
However, the generated features are often over-smoothed, resulting in muffled synthetic speech~\cite{tokuda2013speech}. 
This was a central issue when using HMM~\cite{tokuda2013speech} and also affects more recent DNN-based acoustic models~\cite{ren2022revisiting}. The same issue can be caused by the DSP-based vocoders. Even if the trend is to switch to DNN-based vocoders, as long as the DNN-based vocoders are trained to minimize the distance of the spectral magnitude between generated and natural speech, the mean-seeking training paradigm is known to produce blurry (or over-smoothed) speech.

We simulate this effect by applying a Gaussian filter to \gls{lsf} along the temporal axis, reducing frame-to-frame variation~\cite{merritt2013investigating}. 
The filter standard deviation is empirically set to \num{2}. 
After filtering, \gls{lsf} coefficients are sorted within each frame to preserve ordering constraints. This yields trajectories with reduced temporal variability, approximating oversmoothed speech.
Raw \glspl{lsf} were ineffective for training: we aggregate frame-level features over time into a 13-dimensional vector for \gls{mlp}-based classification.

\subsection{Degraded Phase}
\label{subsubsec:exp_3}
Phase artifacts commonly arise in three cases: 1) when using vocoders that assume a minimum-phase when reconstructing waveforms from spectral magnitude~\cite{morise2016world},  2) when using (typically DNN-based) vocoders that do not explicitly generate the phase~\cite{NEURIPS2020_c5d73680}, and 3) when the DNN cannot accurately generate the phase~\cite{siuzdak2023vocos}.

We model this via controlled lossy phase reconstruction. The waveform is converted to the \gls{stft} domain, retaining magnitude while discarding phase. A minimum-phase signal is reconstructed using cepstral processing by enforcing a causal liftering constraint on the log-magnitude spectrum, followed by inverse \gls{stft}. From the reconstructed waveform, the unwrapped phase representation is extracted and used as input to a Light CNN model~\cite{wuLightCNNDeep2018}.

\subsection{Energy Inconsistencies}
\label{subsubsec:exp_4}
This artifact occurs in both unit selection and SPSS-based TTS systems. In unit selection, it may result from imperfect concatenation, while in SPSS it appears as overly uniform regions or abrupt transitions, including unnaturally clean silences.

We simulate this by selecting a short temporal region (\num{6}–\num{30} frames) and attenuating it by a factor drawn from $[0.0, \sqrt{0.5}]$, producing local energy drops. The waveform is then converted to a power spectrogram via \gls{stft} and log-scaled. We aggregate it into \num{24} scalar descriptors capturing energy statistics and temporal drops, which are input to a \gls{mlp}.

\subsection{Tonal Artifacts}
\label{subsubsec:exp_5}
Narrowband tonal artifacts can occur in neural \gls{tts} systems using DNN-based upsampling modules, such as transposed or subpixel convolutions~\cite{pons2021upsampling}. 
These may introduce spectral replicas and uneven overlap, producing spurious narrowband artifacts unrelated to natural harmonics. These are referred to as tonal artifacts~\cite{pons2021upsampling}.

We simulate this by attenuating the waveform by \num{0.05}, then adding \num{1}–\num{2} narrowband resonances using high-Q band-pass filters (1–8 kHz, $Q \in [30, 80]$, gains $[2, 5]$). 
A nonlinearity increases contrast, followed by normalization, yielding signals dominated by sparse spectral peaks. A spectral fingerprint is extracted following~\cite{afchar2025fourier_explanation}, producing a 1D feature emphasizing these peaks for a \gls{mlp}.

\vspace{0.2em}
Figure~\ref{fig:artifacts-examples} shows representative real-vs-synthetic examples for four artifacts, illustrating the features used by the corresponding experts; the degraded phase is omitted due to limited suitability for 2D visualization.
The five artifacts are assumed to cover many speech synthesis systems in publicly available deepfake datasets~\cite{wang2020asvspoof, jung2024spoofceleb}. 
Many voice conversion (VC) systems relying on statistical models and vocoders, DNN-based or not, may exhibit similar artifacts. A limitation of this approach is that only artifacts known a priori can be modeled by an expert. This is further discussed in \S~\ref{sec:conclusion}.

\section{Experimental Setup}
\label{sec:setup}
This section outlines the data splits (\S~\ref{subsec:data}), training protocol (\S~\ref{subsec:training}), and evaluation metrics (\S~\ref{subsec:metrics}) used in our experiments.

\subsection{Datasets and Partitioning Protocol}
\label{subsec:data}
Our training data consists entirely of real speech drawn from the training sets of ASVspoof 2019~\cite{wang2020asvspoof} (clean studio recordings), ASVspoof 5~\cite{WANG2026101825} (controlled recordings), and SpoofCeleb~\cite{jung2024spoofceleb} (in-the-wild speech). We trim their leading and trailing silence and apply peak normalization to avoid potential shortcut learning during training~\cite{muller21_asvspoof}.
To keep balance, each corpus is subsampled to match the size of the smallest one. 
The resulting subsets are split \num{80}/\num{20} at the speaker level into disjoint training and validation sets, and the splits are concatenated across corpora. 
Every utterance is duplicated for each expert: one copy is kept unchanged, while the other is modified to match the target artifact. 
After duplication, the final dataset contains \num{24526} training samples and \num{6242} validation samples.

For calibration, real samples are drawn from the validation set. 
When paired with actual deepfake samples, these are taken either from ASVspoof 2015 (attack S3 for $F_0$ and energy experts, S10 for the others) or from the ASVspoof 2019 training set (A01 for the tonal artifact expert, A03 for the oversmoothing and phase experts, A04 for the $F_0$ and energy ones), with A02 excluded for redundancy but retained for testing. 
Further details on the choice of these attacks are provided in~\S~\ref{subsec:res_2}.

For evaluation, real samples are taken exclusively from evaluation partitions, while \gls{tts} samples are drawn from the development and evaluation sets for ASVspoof 2019, and from all splits for ASVspoof 5 and SpoofCeleb, which are never used for calibration. 
For ASVspoof 5, post-processed samples are excluded, as robustness assessment falls outside the scope of this initial analysis. 
For both ASVspoof 5 and SpoofCeleb, the number of \gls{tts} samples per attack is clipped to the size of the least represented attack class (\num{7760} and \num{7963} samples, respectively), which remains sufficient for meaningful evaluation while balancing class distributions.

\subsection{Training Specifications}
\label{subsec:training}
Each expert model is trained for up to \num{100} epochs with early stopping after \num{20} epochs without improvement. 
The batch size is \num{128}. 
Optimization uses AdamW with an initial learning rate of $10^{-3}$, reduced on plateau with patience \num{10}. 
Training uses binary cross-entropy, and selection is based on validation loss.
Pseudo-fake samples are generated on-the-fly by applying the corresponding expert's designed manipulation, with dynamic labeling (\num{0} for real, \num{1} for manipulated). 
For \glspl{mlp} using scalar features, inputs are standardized using statistics computed on the training set.

The gating network for fusing LLRs (Eq.~\eqref{eq:gating}) is trained separately for \num{10} epochs under the same setup. 
In addition to binary cross-entropy, it includes an auxiliary negative log-likelihood loss enforcing correct expert-artifact attribution, since this time each pseudo-fake contains a single uniformly sampled manipulation. 

In terms of calibration, the Gaussian \gls{kde} uses a bandwidth of $0.1$. 
For the CMLG approach, the shared variance is experimentally set to $\sigma={0.5}$.

\subsection{Evaluation Metrics}
\label{subsec:metrics}
We report \gls{auc} and \gls{eer} for discriminative performance. \gls{auc} measures ranking quality across thresholds, while \gls{eer} is the operating point where false acceptance and false rejection rates are equal. Following standard practice in speech deepfake detection, validation \gls{eer} is also used for gating network selection.
Both metrics are calibration-blind and depend only on score ordering.\footnote{Any order-preserving transformation of scores leaves \gls{auc} and \gls{eer} unchanged~\cite[\S~3.3]{van2007introduction}.} 
To assess calibration quality, we additionally report \gls{cllr}, which evaluates the quality of estimated \glspl{llr} in terms of both discrimination and calibration~\cite{van2007introduction}. Lower \gls{cllr} indicates better-calibrated and more informative scores, while values above \num{1} indicate poor calibration and weak discrimination.
\section{Results}
\label{sec:results}

\begin{table}
\caption{Experts performance on ASVspoof 2019 attacks exhibiting their target artifacts, reported as AUC ($\uparrow$) and EER ($\downarrow$). Experts were trained using pseudo-fake data. Data from target attacks are unseen.}
\label{tab:target_eval}
\centering
\resizebox{\columnwidth}{!}{
\begin{tabular}{l l c c}
\toprule
\textbf{Expert} & \textbf{Target Attack} & \textbf{AUC (\%)} & \textbf{EER (\%)} \\
\midrule
F0 Inconsistencies 
& A04 (Unit selection)
& 77.64 
& 29.38 \\
Oversmoothing  
& A03 (DNN + WORLD)
& 97.64 
& 6.62 \\
Degraded Phase 
& A03 (DNN + WORLD)
& 98.50 
& 6.24 \\
Energy Inconsistencies 
& A04 (Unit selection)
& 88.23 
& 22.98 \\
Tonal artifacts 
& A01 (DNN + WaveNet)
& 84.77 
& 22.86 \\
\bottomrule
\end{tabular}}
\vspace{-1em}
\end{table}

%%%%%%%%%%%%%%%%%%%%%%%%%%%%%%%%%%%%%%%%%%%%%%%%%%%%%%%%%%%%%%%%%%%%%%%%%%%%%%%%%%%%%%%%%%%%%%%%%%%%%%%%%%%%%%%%%%%%%%%%%%%%%%%%%%%%%%%%%%%%%%%%%%%%%%%%%%%%%%%%%%%%%%%%%%%%%%%%%%%%%%%%%%%%%%%%%%%%%%%%%%%%%%%%%%%%%%%%%%%%%%%%%%%%%%%%%%%%%%%%%%%%%%%%%%%%%%%%%%%%%%%%%%%%%%%%%%%%%%%%%%%%%%%%%%%%%%%%%%%%%%%%%%%%%%%%%%%%%%%%%%%%%%%%%%%%%%%%%%%%%%%%

Experimental results are presented in this section, including the individual evaluation of the artifact-specific experts (\S~\ref{subsec:res_1}), the optimization of calibration data and methods for the ensemble (\S~\ref{subsec:res_2}), a comparative analysis of LLR aggregation strategies (\S~\ref{subsec:res_3}), and an interpretability analysis mapping expert responses to the generation attacks (\S~\ref{subsec:res_4}).

\subsection{Evaluation of Individual Expert}
\label{subsec:res_1}
To showcase the performance of the artifact-specific experts, we measure their detection results on selected deepfake attacks. 
For each expert, we select an ASVspoof 2019 attack that is expected to exhibit the targeted artifact based on its underlying components.
Since we are interested in its discrimination power, each expert's output score $s_i$ is used to compute the AUC and EER without calibration (\S~\ref{subsec:metrics}). 
Results are reported in \Cref{tab:target_eval} alongside the corresponding synthesis architectures. All experts perform above chance level ($\text{AUC}=\text{EER}=50\%$), indicating that the chosen expert–artifact assignments are meaningful. 
Although performance remains well below state-of-the-art detectors, this is expected: each expert uses a compact model and targets a single artifact rather than exploiting all available deepfake cues. 
Plus, experts are trained only on real and pseudo-fake data, not on actual TTS-generated speech (\S~\ref{subsec:met_1}).

\subsection{Evaluation of Calibrated Experts in Ensemble}
\label{subsec:res_2}
\begin{figure*}
    \centering
    \hspace*{2.15cm}
    \includegraphics[width=0.75\textwidth]{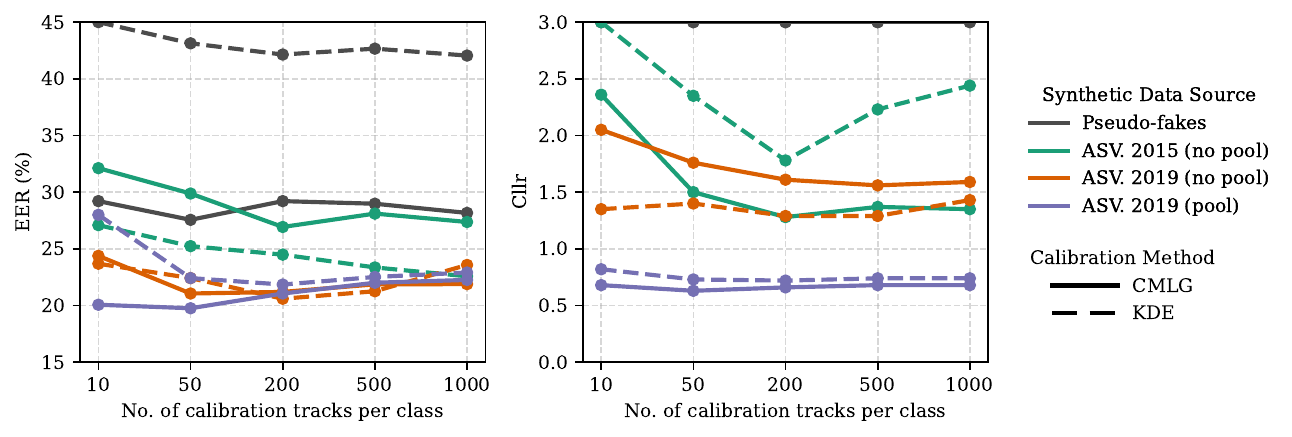}
    \caption{Ensemble overall performance on ASVspoof 2019 for different calibration functions, calibration set compositions, and calibration set sizes: \gls{eer} (left) and \gls{cllr} (right). \gls{cllr} values above 3 are clipped for visualization.}
    \label{fig:calibration_plots}
    \vspace{-1em}
\end{figure*}

Evaluation is first conducted independently for each expert, then outputs are aligned by retaining only samples valid across all experts (discarding those with failed $F_0$, which is \num{3.24}\%).
Performance of the ensemble of experts depends on how well the experts are calibrated and the strategy to fuse the calibrated LLRs. Hence, in this section, 
we first evaluate calibration choices in \Cref{fig:calibration_plots} using a fixed sum aggregation. 
We then compare aggregation strategies under the selected calibration setup in the next subsection.

The goodness of calibration is affected by the calibration data as well as the calibration method. 
In \Cref{fig:calibration_plots}, for the two calibration methods using either CMLG or KDE (\S~\ref{subsec:met_2}), we evaluate \num{4} calibration set compositions with different data sources: 
1) pseudo-fakes as in training; 2) ASVspoof 2015 attacks, where each expert uses only a selected attack that matches its target artifact (see \S~\ref{subsec:data}); 3) ASVspoof 2019 attacks under the same expert-specific setup; and 4) a pooled ASVspoof 2019 variant, where the target attacks from all experts are merged into a shared calibration set used for every expert.
Each calibration set is further prepared in \num{5} sizes, ranging from \num{10} to \num{1000} fake utterances per attack.
The real utterances for each set, drawn from validation set, are subsampled to be equal to the number of fake samples.
The evaluation set is ASVspoof 2019.

We observe that only the 4th choice (i.e., the pooled ASVspoof 2019 variant) yields acceptable \gls{cllr} values (below \num{1.0}), indicating that a shared calibration set across experts leads to a more stable ensemble, with the best results obtained using the CMLG calibration function. 
In this condition, the number of utterances is a negligible factor; the gap between KDE and the CMLG-based calibration methods is also small. 
Specifically, the best setup scores \num{19.76}\% \gls{eer} and \num{0.638} \gls{cllr}, using only \num{50} samples per calibration class (i.e., each attack type).

\subsection{Evaluation of Experts Aggregation}
\label{subsec:res_3}
Using the best calibration data and method identified in the previous subsection, we compared three aggregation methods (\S~\ref{subsec:met_3}) on the ASVspoof 2019 evaluation set. The results presented in \Cref{tab:aggregation} show that summing expert \glspl{llr} consistently achieves the best performance in terms of both \gls{eer} and Cllr, outperforming max-based and learned gating strategies (\Cref{tab:aggregation}). 

\begin{table}
\caption{Ensemble aggregation strategies comparison on ASVspoof 2019 evaluation set. EER ($\downarrow$), Cllr ($\downarrow$).}
\label{tab:aggregation}
\centering
\begin{tabular}{l c c}
\toprule
\textbf{Aggregation Method} & \textbf{EER (\%)} & \textbf{Cllr} \\
\midrule
Sum of Expert LLRs      & \textbf{19.76} &\textbf{0.638} \\
Maximum Expert LLR      & 22.22 & 0.855 \\
Gating Network          & 21.54 & 0.907 \\
\bottomrule
\end{tabular}
\vspace{-1em}
\end{table}

%%%%%%%%%%%%%%%%%%%%%%%%%%%%%%%%%%%%%%%%%%%%%%%%%%%%%%%%%%%%%%%%%%%%%%%%%%%%%%%%%%%%%%%%%%%%%%%%%%%%%%%%%%%%%%%%%%%%%%%%%%%%%%%%%%%%%%%%%%%%%%%%%%%%%%%%%%%%%%%%%%%%%%%%%%%%%%%%%%%%%%%%%%%%%%%%%%%%%%%%%%%%%%%%%%%%%%%%%%%%%%%%%%%%%%%%%%%%%%%%%%%%%%%%%%%%%%%%%%%%%%%%%%%%%%%%%%%%%%%%%%%%%%%%%%%%%%%%%%%%%%%%%%%%%%%%%%%%%%%%%%%%%%%%%%%%%%%%%%%%%%%%

\subsection{Interpretability Analysis}
\label{subsec:res_4}
In \Cref{tab:artifact_asv19}, we report the \glspl{llr} of the experts in the best-performing ensemble identified in \S~\ref{subsec:res_2}, computed for each ASVspoof 2019 attack.
Overall, average \glspl{llr} remain lower for real than fake samples across all experts, indicating consistent ensemble behavior. 
Also, the expert responses are consistent with the characteristics of the underlying generation pipeline of most deepfake attacks: for example, attacks based on HMM/DNN acoustic models consistently yield positive \glspl{llr} from the oversmoothing expert, while attacks employing DSP-based (WORLD, Vocaine) or DNN vocoders without phase conditioning (Neural Source-Filter) systematically activate the phase expert.
This behavior is expected, as the experts are designed to capture specific types of artifacts, which match the generation techniques represented in the dataset used. 
Moreover, the strength of the expert responses varies considerably across attacks. 
Some attacks elicit only weak activations, whereas others trigger multiple experts simultaneously, indicating the coexistence of several artifacts within the same generation pipeline.
In contrast to the other experts, the $F_0$ one exhibits weaker responses across most attacks, consistent with its lower pre-calibration discriminative performance (\Cref{tab:target_eval}) and suggesting that $F_0$-related artifacts might be less consistently present or expressed in these attacks.

We then examine the performance of experts on ASVspoof 5 and SpoofCeleb, which include newer-generation attacks. As long as the artifacts persist in the newer attacks, they are expected to be spotted by the experts.
The calibration set is kept unchanged from the ASVspoof 2019 setup.

On ASVspoof 5 (\Cref{tab:artifact_asv5}), oversmoothing and degraded phase remain consistently activated across many attacks, suggesting partial persistence of these artifacts in newer \gls{tts} systems. 
In contrast, energy inconsistencies largely disappear, with only a few isolated exceptions (Tacotron2+WaveGrad exhibits unnaturally blank silences). 
A similar pattern holds for tonal artifacts, which remain active only for specific systems and partially overlap with those triggering the energy expert, with responses concentrated mainly on Toucan TTS-based attacks across different vocoders.

SpoofCeleb (\Cref{tab:artifact_sc}) is overall more challenging but also shows promising results. 
Oversmoothing and degraded phase remain the most prominent artifacts, while energy inconsistencies almost vanish.
The tonal expert responds strongly only to WaveGlow-based systems, consistent with their characteristic audible distortions.
Notably, real samples are not always assigned negative \glspl{llr} across experts, suggesting artifact-like characteristics in in-the-wild recordings. 
While not desirable, this is not unexpected, as expert outputs represent evidence of specific artifacts rather than global authenticity. 
Global decisions are instead obtained by aggregating these signals at the ensemble level into $H_0$ (real) and $H_1$ (fake), which is meant to help mitigate inconsistencies across individual experts.

Overall, the interpretability analysis shows that the artifact-specific experts yield expected results on ASVspoof 2019 by producing positive LLRs to their targeted synthesis pipelines. Furthermore, evaluation on newer datasets demonstrates that these experts also generalize well to varied extents, catching persistent acoustic and phase distortions in modern architectures.

\begin{table*}
\caption{Expert \glspl{llr} on ASVspoof 2019. Higher values indicate stronger evidence in favor of the fake hypothesis ($H_1$). \textbf{Bold} values indicate positive \glspl{llr}.}
\label{tab:all_artifact_asv19}
\vspace{-1em}
\begin{subtable}{\textwidth}
\caption{Artifact-specific experts reporting calibrated LLRs per attack. The first three attacks are included in each expert’s calibration set.}
\label{tab:artifact_asv19}
\resizebox{\textwidth}{!}{
\begin{tabular}{l|c|ccc|ccccccc|c}
\hline
\toprule
& \textbf{Avg Real} 
& \shortstack{HMM/DNN\\WaveNet}
& \shortstack{DNN\\WORLD}
& \shortstack{Unit Selection}
& \shortstack{HMM/DNN\\WORLD}
& \shortstack{DNN\\WORLD+GAN}
& \shortstack{HMM/DNN\\Neural SF}
& \shortstack{DNN\\Vocaine}
& \shortstack{Tacotron2\\WaveRNN}
& \shortstack{Tacotron2\\Griffin-Lim}
& \shortstack{DNN\\WaveNet}
& \textbf{Avg Fake} \\
\midrule \midrule
F0 Inconsistencies      & 0.00  & -0.01 & \textbf{0.02} & -0.02 & \textbf{0.03} & \textbf{0.03} & 0.00 & \textbf{0.03} & 0.00 & \textbf{0.01} & 0.00 & \textbf{0.01} \\
Oversmoothing           & -0.10 & \textbf{0.05}  & \textbf{0.30} & -0.20 & \textbf{0.19} & -0.18 & \textbf{0.13} & \textbf{0.24} & -0.16 & \textbf{0.08} & -0.25 & \textbf{0.02} \\
Degraded Phase          & -0.39 & \textbf{0.04}  & \textbf{0.71} & -0.37 & -0.17 & \textbf{0.27} & \textbf{0.61} & \textbf{1.03} & \textbf{0.06} & \textbf{0.20} & -0.10 & \textbf{0.23} \\
Energy Inconsistencies  & -0.49 & -0.77 & \textbf{0.63} & \textbf{1.77} & -0.44 & -0.32 & \textbf{0.01} & \textbf{1.25} & -0.28 & \textbf{1.52} & \textbf{2.09} & \textbf{0.55} \\
Tonal Artifacts         & -0.09 & \textbf{0.60}  & \textbf{0.31} & -0.08 & \textbf{0.17} & \textbf{0.06} & \textbf{0.44} & \textbf{0.04} & -0.10 & -0.05 & -0.11 & \textbf{0.13} \\
\bottomrule
\end{tabular}
}
\end{subtable}

\vspace{0.5em}

\begin{subtable}{\textwidth}
\caption{Additional silence duration expert reporting per-attack calibrated LLRs. The first three attacks are part of its calibration set.}
\label{tab:asv19_add_exp}
\resizebox{\textwidth}{!}{
\begin{tabular}{l|c|ccc|ccccccc|c}
\hline
\toprule
& \textbf{Avg Real} 
& \shortstack{HMM/DNN\\WaveNet}
& \shortstack{DNN\\WORLD}
& \shortstack{Unit Selection}
& \shortstack{HMM/DNN\\WORLD}
& \shortstack{DNN\\WORLD+GAN}
& \shortstack{HMM/DNN\\Neural SF}
& \shortstack{DNN\\Vocaine}
& \shortstack{Tacotron2\\WaveRNN}
& \shortstack{Tacotron2\\Griffin-Lim}
& \shortstack{DNN\\WaveNet}
& \textbf{Avg Fake} \\
\midrule \midrule
Silence Duration Bias  & -0.19  & \textbf{0.48} & -0.08 & \textbf{0.30} & -0.13 & \textbf{1.88} & -0.03 & \textbf{0.91} & \textbf{2.27} & \textbf{2.05} & \textbf{0.59} & \textbf{0.82} \\
\bottomrule
\end{tabular}
}
\end{subtable}

\end{table*}

\begin{table*}
\caption{Expert \glspl{llr} on ASVspoof 5, reported per attack. Top: acoustic model; bottom: vocoder or no distinction. Higher LLRs indicate stronger evidence in favor of the fake hypothesis ($H_1$). \textbf{Bold} values indicate positive LLRs.}
\vspace{-1em}
\label{tab:artifact_asv5}
\setlength{\tabcolsep}{3pt}
\resizebox{\textwidth}{!}{
\begin{tabular}{l|c|ccccccccccccccc|c}
\hline
\toprule
& \textbf{Avg Real}
& \shortstack{GlowTTS\\HiFiGAN}
& \shortstack{GradTTS\\HiFiGAN}
& \shortstack{VAE\\HiFiGAN}
& \shortstack{XTTS\\HiFiGAN}
& \shortstack{VITS\\HiFiGAN}
& \shortstack{YourTTS}
& \shortstack{YourTTS v2}
& \shortstack{FastPitch\\HiFiGAN}
& \shortstack{Toucan TTS\\HiFiGAN}
& \shortstack{Toucan TTS\\HiFiGAN v2}
& \shortstack{Toucan TTS\\BigVGAN}
& \shortstack{Toucan TTS\\BigVGAN v2}
& \shortstack{Tacotron2\\WaveGrad}
& \shortstack{Unit Sel.}
& \shortstack{Unit Sel.}
& \textbf{Avg Fake} \\
\midrule \midrule
F0 Inconsistencies   & -0.01 & 0.00 & 0.00 & \textbf{0.01} & \textbf{0.01} & \textbf{0.02} & -0.01 & \textbf{0.01} & \textbf{0.02} & \textbf{0.02} & \textbf{0.02} & \textbf{0.02} & \textbf{0.01} & \textbf{0.01} & -0.02 & \textbf{0.01} & \textbf{0.01} \\
Oversmoothing        & -0.16 & \textbf{0.30} & \textbf{0.24} & \textbf{0.11} & \textbf{0.18} & \textbf{0.21} & \textbf{0.08} & \textbf{0.06} & \textbf{0.46} & \textbf{0.23} & \textbf{0.23} & \textbf{0.30} & \textbf{0.20} & \textbf{0.13} & -0.33 & \textbf{0.04} & \textbf{0.15} \\
Degraded Phase       & -0.14 & \textbf{0.17} & \textbf{0.24} & \textbf{0.06} & \textbf{0.30} & \textbf{0.05} & \textbf{0.23} & \textbf{0.18} & \textbf{0.35} & \textbf{0.26} & \textbf{0.26} & \textbf{0.17} & \textbf{0.29} & \textbf{0.03} & -0.36 & \textbf{0.06} & \textbf{0.14} \\
Energy Inconsist.    & -0.25 & -0.75 & -0.70 & -0.57 & -0.57 & -0.77 & -0.89 & -0.41 & -0.81 & \textbf{0.14} & \textbf{0.15} & \textbf{0.12} & \textbf{0.61} & \textbf{2.06} & -0.17 & -0.57 & -0.19 \\
Tonal Artifacts      & -0.13 & -0.22 & -0.28 & -0.18 & -0.24 & -0.13 & -0.35 & -0.13 & -0.29 & \textbf{0.17} & \textbf{0.16} & \textbf{0.01} & \textbf{0.18} & -0.10 & -0.13 & -0.06 & -0.11 \\
\bottomrule
\end{tabular}}
\end{table*}

\begin{table*}
\caption{Expert \glspl{llr} on SpoofCeleb, reported per attack. Top: acoustic model; bottom: vocoder or no distinction. Higher LLRs indicate stronger evidence in favor of the fake hypothesis ($H_1$). \textbf{Bold} values indicate positive LLRs.}
\vspace{-1em}
\label{tab:artifact_sc}
\setlength{\tabcolsep}{2pt}
\resizebox{\textwidth}{!}{
\begin{tabular}{l|c|cccccccccccccccccc|c}
\hline
\toprule
& \textbf{Avg Real}
& \shortstack{VITS}
& \shortstack{MQTTS}
& \shortstack{VALL-E}
& \shortstack{Delay}
& \shortstack{GradTTS\\DiffWave}
& \shortstack{MatchaTTS\\DiffWave}
& \shortstack{GradTTS\\BigVGAN}
& \shortstack{MatchaTTS\\BigVGAN}
& \shortstack{GradTTS\\WaveGlow}
& \shortstack{MatchaTTS\\WaveGlow}
& \shortstack{GradTTS\\NSF HiFiGAN}
& \shortstack{MatchaTTS\\NSF HiFiGAN}
& \shortstack{BVAE-TTS\\NSF HiFiGAN}
& \shortstack{GradTTS\\HiFiGAN}
& \shortstack{MatchaTTS\\HiFiGAN}
& \shortstack{BVAE-TTS\\HiFiGAN}
& \shortstack{MS-Transf}
& \shortstack{TransfTTS\\PWG}
& \textbf{Avg Fake} \\
\midrule \midrule
F0 Inconsistencies   & \textbf{0.01} & 0.00 & \textbf{0.01} & \textbf{0.01} & \textbf{0.01} & 0.00 & \textbf{0.01} & 0.00 & \textbf{0.01} & -0.02 & -0.02 & \textbf{0.01} & \textbf{0.01} & \textbf{0.01} & \textbf{0.01} & \textbf{0.01} & \textbf{0.01} & \textbf{0.01} & \textbf{0.01} & \textbf{0.01} \\
Oversmoothing        & \textbf{0.31} & \textbf{0.14} & \textbf{0.25} & \textbf{0.24} & \textbf{0.22} & \textbf{0.05} & \textbf{0.16} & \textbf{0.22} & \textbf{0.26} & -0.04 & \textbf{0.29} & \textbf{0.33} & \textbf{0.32} & \textbf{0.30} & \textbf{0.33} & \textbf{0.33} & \textbf{0.35} & \textbf{0.31} & \textbf{0.32} & \textbf{0.24} \\
Degraded Phase       & -0.06 & \textbf{0.24} & \textbf{0.25} & \textbf{0.07} & \textbf{0.32} & 0.00 & \textbf{0.24} & -0.22 & \textbf{0.05} & -0.05 & -0.13 & \textbf{0.44} & -0.06 & -0.20 & -0.06 & \textbf{0.42} & -0.03 & -0.05 & -0.05 & \textbf{0.07} \\
Energy Inconsist.    & -1.10 & -0.74 & -0.41 & -0.86 & -0.79 & -1.16 & -0.98 & -0.97 & -0.86 & -1.03 & -0.95 & -0.94 & -1.10 & -1.09 & -1.10 & -0.94 & -1.10 & -1.10 & -1.10 & -0.96 \\
Tonal Artifacts      & -0.36 & \textbf{0.18} & -0.08 & -0.08 & -0.01 & \textbf{0.08} & \textbf{0.10} & -0.01 & -0.05 & \textbf{3.01} & \textbf{3.02} & -0.11 & -0.36 & -0.35 & -0.35 & -0.10 & -0.34 & -0.36 & -0.35 & \textbf{0.21} \\
\bottomrule
\end{tabular}}
\end{table*}

\subsection{Interpretability for Shortcut Diagnosis}
\label{subsec:res_5}
All experts in this work operate on features treated as nuisance-robust representations and are not intended to exploit any dataset-specific shortcuts~\cite{sahidullah2025shortcut}, but we can add experts to diagnose potential shortcuts. 
The impact of a shortcut can be revealed by comparing 1) the values of calibrated \glspl{llr} from the suspected expert with others and 2) the ensemble's performance with and without the suspected expert. The latter can be easily done since the framework is inherently modular: expert outputs are combined as a linear sum, and an expert can be directly added to or removed from the ensemble. No modification or retraining is required for other experts.

As a demonstration, we introduce a 6th control expert targeting a known dataset bias related to silence-duration patterns~\cite{muller21_asvspoof}.
Silence intervals are extracted by applying a voice activity detector and deriving complementary non-speech regions. 
To generate pseudofakes, the silence-duration sequence is stochastically compressed toward its mean, perturbed with controlled noise, truncated for overly long pauses, and constrained at sequence boundaries. 
This reflects the tendency of synthetic speech to exhibit reduced variability and shorter silences, particularly at leading and trailing positions~\cite{muller21_asvspoof}. 
Features are summarized into \num{10} proxies and classified using a dedicated MLP (10–32–16–1).
This expert is designed, trained, and calibrated independently from the others. 
Its calibration set differs slightly from those of the other experts, as real samples are drawn exclusively from ASVspoof 2019 due to the dataset-dependent nature of the targeted artifact.

\begin{figure}
\centering
\includegraphics[width=\columnwidth]{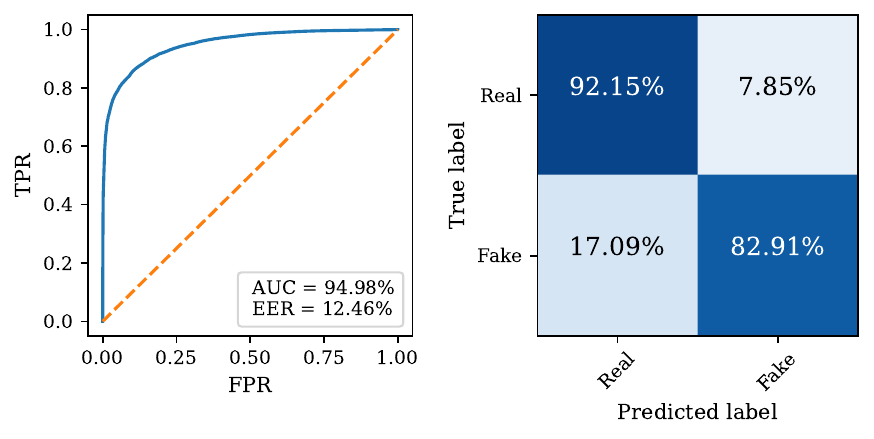}
\caption{Ensemble overall performance on ASVspoof 2019 after adding the sixth expert. ROC curve and confusion matrix (LLR scores threshold set at 0).}
\label{fig:ensemble_6_exp}
\vspace{-1em}
\end{figure}

Results on ASVspoof 2019 \gls{tts} attacks in \Cref{tab:asv19_add_exp} show expected traits of shortcut: particularly for Tacotron2 + WaveRNN, where previous experts were weak, the added expert provides very strong `evidence' (\gls{llr} > \num{2}). At the ensemble level, the expert reduced the EER to around 12\%. However, since the expert only relies on silence duration, the large magnitude of \gls{llr} and reduced EER suggest that the expert is exploiting the dataset-specific shortcut.
We show the new ensemble performance in \Cref{fig:ensemble_6_exp}, along with the confusion matrix at threshold \num{0}.

\section{Conclusions}
\label{sec:conclusion}
We proposed an interpretable speech deepfake detection framework based on artifact-specific expert models. The system decomposes detection into learning and recognition of predefined synthesis artifacts, each modeled by a dedicated classifier trained on pseudo-fake data.
Expert outputs are calibrated into \glspl{llr}, providing a common interpretable measure of evidence linking decisions to artifact presence.
Experiments show that (i) experts reliably detect their target artifacts, (ii) calibration is essential for meaningful comparison across experts, and (iii) \gls{llr} summation is the most effective aggregation strategy.
The framework provides an interpretable mechanism for identifying synthesis artifacts while maintaining a modular architecture that can be extended as new artifact classes emerge. It also helps to diagnose potential dataset shortcuts.

Future work will focus on broadening the range of artifact-specific experts to capture the increasingly diverse characteristics of modern speech synthesis systems. Adapting automated rule-learning frameworks from the visual domain~\cite{raza2024ruleboost} to extract and compose speech anomalies into human-interpretable artifacts represents a promising direction for long-term, extendable development. 
Other topics, such as improved calibration and refined expert pipelines, are also expected to further boost the overall performance.
Owing to its additive and modular design, these developments can be incorporated by introducing new experts while preserving the interpretability and extensibility of the overall framework.

\begin{acks}
This work is partially supported by JST, PRESTO Grant Number JPMJPR23P9, Japan.
\end{acks}

\bibliographystyle{ACM-Reference-Format}
\bibliography{bibliography}

\end{document}